\documentclass[12pt]{article}
\usepackage{graphicx}

\def\leaderfill{\leaders\hbox to 1em{\hss.\hss}\hfill}  

\begin{document}

\noindent{\small\it  ISSN 1063-7729, Astronomy Reports, Vol. 52,
No. 1, 2008, pp. 951--962. \copyright Pleiades Publishing, Ltd.,
2008. \noindent Original Russian Text \copyright A.T. Bajkova,
2008, published in Astronomicheski$\check{\imath}$ Zhurnal, 2008,
Vol. 85, No. 12, pp. 1059--1071. }

\vskip -4mm

\begin{tabular}{lllllllllllllllllllllllllllllllllllllllllll}
 & & & & & & & & & & & & & & & & & & & & & & & & & & & & & & & & & &  \\
\hline \hline
\end{tabular}

\vskip 1.5cm

\bigskip

\centerline{\bf\large Multi-Frequency Synthesis of VLBI Images
Using a Generalized} \centerline{\bf\large Maximum Entropy Method
}

\bigskip

\centerline{\bf A. T. Bajkova}

\bigskip

\centerline{\it Main Astronomical Observatory, Russian Academy of
Sciences} \centerline{\it Pulkovo, St.Petersburg, 196140 Russia}
\centerline{\small Received February 26, 2008; in final form, July
2, 2008}

\bigskip

{\bf Abstract} -- A new multi-frequency synthesis algorithm for
reconstructing images from multi-frequency VLBI data is proposed.
The algorithm is based on a generalized maximum-entropy method,
and makes it possible to derive an effective spectral correction
for images over a broad frequency bandwidth, while simultaneously
reconstructing the spectral-index distribution over the source.
The results of numerical simulations demonstrating the
capabilities of the algorithm are presented.

\medskip


\medskip

{\bf DOI}: 10.1134/S1063772908120019

\vskip 1 cm

\centerline{1.~INTRODUCTION}

\medskip

Coverage of the $UV$ spatial-frequency plane in VLBI observations
[1] is mainly achieved using(1) several antennas, (2) the diurnal
rotation of the Earth, (3) relocation of the antennas. However,
the interferometer aperture is never filled completely due to the
limited observation time and number of antennas, interruptions
during the observations, and limitations on the smallest and
longest baselines. Since the baseline coordinates $U$ and $V$ in
the visibility function are measured in wavelengths, an
alternative method of filling the $UV$ plane is to observe a
source at several frequencies simultaneously.

This method for fast filling of the aperture is called
multi-frequency synthesis (MFS), and its effect is greatest in the
case of interferometers with relatively few baselines. Figure 1a
shows an example of a poor filling of the $UV$ plane obtained by a
four-element interferometer during12 h of observations of a radio
source with a declination of 34. at 8.2 GHz. The coordinates of
the four telescopes were taken to be those of the three stations
in the Russian QUASAR VLBI complex Svetloe, Zelenchukskaya, and
Badary [2], together with the Matera station. Figures 1b, 1c, and
1d present the filling of the $UV$ plane for MFS observations of
this same source with this interferometer using various frequency
ranges. Since the coordinates of the interferometer baseline are
proportional to the frequency, different frequencies within the
total band refer to different, radially arranged $(U, V)$ points.
As we can see, MFS can provide a much denser filling of the $UV$
plane, as well as increase the diameter of the synthesized
aperture compared to the single frequency observations.

However, the application of MFS is complicated by the fact that
the source brightness is a function of frequency; to avoid imaging
artifacts due to this frequency dependence, we must derive and
apply a spectral correction for the maps as they are deconvolved
(reconstructed).

This problem has been considered by many authors. The most
fundamental results on spectral effects in images and methods for
correcting them are presented in [3–7]. These show that, when
sufficiently narrow frequency bands are used, with widths within
$\pm 12.5\%$ of the reference frequency, we can essentially
disregard the spectral dependence of the source brightness, since
these effects are small and can be corrected for during
calibration or self-calibration. However, for broader bands, e.g.,
$\pm 30\%$, correction for spectral effects is necessary.

The most studied linear spectral-correction algorithm is based on
the CLEAN method [8], and is called "double deconvolution" [3].
This algorithm involves the successive deconvolution of a "dirty"
image using a "dirty" beam and using the beam corresponding to the
spectral term. The development of this method enabling the
simultaneous reconstruction of the desired image and the
corresponding spectral map was proposed in [5]. The
vector-relaxation algorithm [9] represents a generalized CLEAN
deconvolution that can take into account spectral terms of any
order.

An alternative deconvolution method that is also actively used in
radio astronomy is the maximum entropy method (MEM). MEM was first
proposed in [10] and [11] for the reconstruction of images in the
optical and radio, respectively. Since then the method has been
developed in many studies [12– 18] and implemented in a number of
software packages designed for image reconstruction (MEMSYS, AIPS,
VLBImager, etc.). A comparative analysis of CLEAN and MEM in VLBI
is given in [16]; essentially, these methods complement each
other. In particular, CLEAN is preferential for reconstructing
images of compact sources from relatively poor data, while MEM is
more suitable for imaging extended sources from better-quality
data. A severe drawback of MEM compared to CLEAN, the bias of the
solution [16], can easily be removed by generalizing the method to
enable the reconstruction of sign-variable functions [17, 18].
This generalized MEM also enables difference imaging making it
possible to substantially broaden the dynamic range of maps of
sources including both compact and extended, faint components
[19]. An important technical advantage of MEM over CLEAN is the
absence of the need to discretize (grid) [1] the visibility
function: MEM does not involve calculation of the inverse Fourier
transform of the visibility data.

The purpose of the current study was to realize in an effective
MFS aperture-synthesis method all the advantages of MEM, in order
to enable high precision correction for spectral effects in a
broad frequency band, as well as to study the developed algorithm
for broad ranges of the various parameters involved.

\bigskip

\centerline{2.~SPECTRAL DEPENDENCE}

\bigskip

The dependence of the source intensity on the frequency í for
synchrotron emission is usually described using the relation [3–7]

\begin{equation}
I(\nu)=I(\nu_0)\left(\frac{\nu}{\nu_0}\right)^\alpha,
\end{equation}
where $I(\nu_0)$ is the intensity at the reference frequency í0
and á is the spectral index. For simplicity, we will use the
notation $I_0=I(\nu_0)$.

Restricting our consideration to the first $Q$ terms of the Taylor
expansion (1) at the point $\nu_0$, we can write the approximate
equality

\begin{equation}
I(\nu)\approx I_0+\sum_{q=1}^{Q-1}I_q
\left(\frac{\nu-\nu_0}{\nu_0}\right)^q,
\end{equation}
where
$$I_q=I_0\frac{\alpha(\alpha-1)\cdots[\alpha-(q-1)]}{q!}.$$

In accordance with (2), we have for each point $(l,m)$ in the
two-dimensional ($N\times N$) brightness distribution (image, map)
of the source

\begin{equation}
I(l,m)\approx I_0(l,m)+\sum_{q=1}^{Q-1}I_q(l,m)
\left(\frac{\nu-\nu_0}{\nu_0}\right)^q,
\end{equation}
where $l,m=1,...,N$.

Thus, the obtained brightness distribution of the source (3)
represents the sum of the brightness distribution at the reference
frequency $\nu_0$ and various spectral terms. The $q$th-order
spectral map depends on the spectral-index distribution in the
source as

\begin{equation}
I_q(l,m)=I_0(l,m)\frac{\alpha(l,m)[\alpha(l,m)-1]\cdots[\alpha(l,m)-(q-1)]}{q!}.
\end{equation}

The first-order spectral map,

\begin{equation}
I_1(l,m)=I_0(l,m)\alpha(l,m)
\end{equation}
is of primary interest.

We can obtain from (5) the estimate of the spectral index
distribution in the source
\begin{equation}
\alpha(l,m)=I_1(l,m)/I_0(l,m).
\end{equation}

\bigskip

\centerline{3.~RESTRICTIONS ON THE VISIBILITY FUNCTION}

\bigskip

The complex visibility function represents the Fourier transform
of the source intensity distribution, which satisfies the spectral
dependence (1) at each point $(l,m)$ of the map. In view of the
finite order of the Taylor series expansion (3), we can write the
restrictions on the visibility function as
\begin{equation}
V_{u_\nu,v_\nu}=  {\bf F}\{I(l,m)\}\times {\bf
D}_{u_\nu,v_\nu}\approx \sum_{q=0}^{Q-1} {\bf F}\left\{I_q(l,m)
\left(\frac{\nu-\nu_0}{\nu_0}\right)^q\right\}\times {\bf
D}_{u_\nu,v_\nu},
\end{equation}
where {\bf F} denotes the Fourier transform and D is the transfer
function, which is a $\delta$ function of $u$ and $v$ for each
visibility measurement; each used frequency í corresponds to its
own set of $\delta$ functions, indicated by the subscripts of $u$
and $v$.

Let us rewrite (7) for the real and imaginary parts of the
visibility function
$V_{u_\nu,v_\nu}=A_{u_\nu,v_\nu}+jB_{u_\nu,v_\nu}$ taking into
account the measurement errors:

\begin{equation}
\sum_{q=0}^{Q-1}\sum_{l,m} I_q(l,m)
a^{lm}_{u_\nu,v_\nu}\left(\frac{\nu-\nu_0}{\nu_0}\right)^q+\eta^{re}_{u_\nu,v_\nu}=A_{u_\nu,v_\nu},
\end{equation}

\begin{equation}
\sum_{q=0}^{Q-1}\sum_{l,m} I_q(l,m)
b^{lm}_{u_\nu,v_\nu}\left(\frac{\nu-\nu_0}{\nu_0}\right)^q+\eta^{im}_{u_\nu,v_\nu}=B_{u_\nu,v_\nu},
\end{equation}
where $a^{lm}_{u_\nu,v_\nu}$ and $b^{lm}_{u_\nu,v_\nu}$ are
constant factors (cosines and sines) that correspond to the
Fourier transform; $\eta^{re}_{u_\nu,v_\nu}$ and
$b^{lm}_{u_\nu,v_\nu}$ are the real and imaginary parts of the
additive instrumental noise, which has a normal distribution with
zero mean and a known dispersion $\sigma_{u_\nu,v_\nu}$.

\bigskip

\centerline{4.~RECONSTRUCTION METHOD}

\bigskip

In our case, the unknowns are the distributions $I_q(l,m)$
($q=0,...,Q-1;~~ l,m=1,...,N$) and the errors of the visibility
function $\eta^{re}_{u_\nu,v_\nu},~\eta^{im}_{u_\nu,v_\nu}$. Note
that, in spite of the fact that the source brightness distribution
is described by a nonnegative function, in general, spectral maps
of arbitrary order (4) can take on both positive and negative
values, due to the sign-variable spectral-index distribution in
the source. Since the logarithm of a negative value is not defined
on the set of real numbers, we will compose for the purpose of
finding a solution for $I_q(l,m),~q=0,...,Q-1$ a functional in
which the values of the spectral maps are represented by their
absolute values:
\begin{equation}
{\bf E}=\{\sum_{l,m} I_0(l,m)\ln [I_0(l,m)]+\sum_{q=1}^{Q-1}
\sum_{l,m} |I_q(l,m)|\ln [|I_q(l,m)|]\}
+\rho\sum_{u_\nu,v_\nu}\frac{(\eta^{re}_{u_\nu,v_\nu})^2+(\eta^{im}_{u_\nu,v_\nu})^2}{\sigma^2_{u_\nu,v_\nu}},
\end{equation}
\begin{equation}
I_0(l,m)\ge 0,
\end{equation}
where $\rho$ is a positive weighting coefficient.

The minimization of the functional (10) with the restrictions (8),
(9), and (11) is the essence of MEM.

We can see from (10) that the minimized functional has two parts:
the entropy functional in the Shannon form and a term representing
an estimate of the ÷2 residual for the difference between the
reconstructed spectrum and the data. This term can be considered
an additional regularizing or stabilizing term that facilitates
further regularization of the solution beyond that possible with
the entropy functional alone [18]. We will bear in mind the effect
of this term on the resolving power of the reconstruction
algorithm.

For carrying out the reconstruction in practice, we now proceed to
the generalized maximum-entropy method described in detail in
[17–19]. This consists in the substitution
\begin{equation}
I_q(l,m)=I_q^+(l,m)-I_q^-(l,m),
\end{equation}
where the superscripts $+$ and $-$ denote the positive and
negative parts of the function, and, in the following modification
of functional (10),
\begin{eqnarray}
{\bf E}=\sum_{l,m} I_0(l,m)\ln [a I_0(l,m)]+\sum_{q=1}^{Q-1}
\sum_{l,m}  \{I_q^+(l,m)\ln [a I_q^+(l,m)]+ \nonumber
 \\+I_q^-(l,m)\ln [a I_q^-(l,m)] \}
+\rho\sum_{u_\nu,v_\nu}\frac{(\eta^{re}_{u_\nu,v_\nu})^2+(\eta^{im}_{u_\nu,v_\nu})^2}{\sigma^2_{u_\nu,v_\nu}},
\end{eqnarray}

\begin{equation}
I_0(l,m)\ge 0,~~~~~I_q^+(l,m)\ge 0,~~~~~I_q^-(l,m)\ge 0,
\end{equation}
where $a\gg 1$ is a parameter responsible [as will be shown below,
see (24)] for the accuracy with which the positive $I_q^+(l,m)$
and negative $I_q^-(l,m)$ parts of the solution for $I_q(l,m)$ can
be separated [see (12)].

The linear restrictions (8) and (9) on the visibility data are
accordingly rewritten
\begin{equation}
{\bf R_A}=\sum_{l,m} I_0(l,m)
a^{lm}_{u_\nu,v_\nu}+\sum_{q=1}^{Q-1}\sum_{l,m}
[I_q^+(l,m)-I_q^-(l,m)]
a^{lm}_{u_\nu,v_\nu}\left(\frac{\nu-\nu_0}{\nu_0}\right)^q+\eta^{re}_{u_\nu,v_\nu}=A_{u_\nu,v_\nu},
\end{equation}
\begin{equation}
{\bf R_B}=\sum_{l,m} I_0(l,m)
b^{lm}_{u_\nu,v_\nu}+\sum_{q=1}^{Q-1}\sum_{l,m}
[I_q^+(l,m)-I_q^-(l,m)]
b^{lm}_{u_\nu,v_\nu}\left(\frac{\nu-\nu_0}{\nu_0}\right)^q+\eta^{im}_{u_\nu,v_\nu}=B_{u_\nu,v_\nu},
\end{equation}
where ${\bf R_A}$ and ${\bf R_B}$ denote the left-hand sides of
the equations.

Thus, the reconstruction of the image I0(l,m) requires the
optimization of the functional (13); i.e., we must find
\begin{equation}
\min {\bf E}
\end{equation}
with the constraints (14)–(16) with respect to all the unknowns
$I_0(l,m),~I_q^{+(-)}(l,m),~q=1,...,Q-1,~l,m=1,...,N $, and
$\eta^{re}_{u_\nu,v_\nu},~\eta^{im}_{u_\nu,v_\nu}$. Note that the
requirements (14) can be omitted since the entropy solution can
only be positive; there then remain only the linear constraints
(15) and (16) on the measurements of the complex visibility
function.

\bigskip

\centerline{6.~THE OPTIMIZATION}

\bigskip

The numerical solution for the optimization of (17) with the
constraints (15) and (16) is based on the method of Lagrange
multipliers; in this method, the conditional optimization is
reduced to an unconditional optimization by composing the
following dual functional, which is called the Lagrange
functional:

\begin{equation}
{\bf L}={\bf E}+\sum_{u_\nu,v_\nu}\{\beta_{u_\nu,v_\nu}({\bf
R_A}-A_{u_\nu,v_\nu}) +\gamma_{u_\nu,v_\nu}({\bf
R_B}-B_{u_\nu,v_\nu})\},
\end{equation}
where $\beta_{u_\nu,v_\nu}$ and $\gamma_{u_\nu,v_\nu}$ are
Lagrange multipliers, or the dual variables with which the
constraints (15) and (16) enter the considered functional.

Thus, the reconstruction is reduced to minimizing the Lag rang e
functional:

\begin{equation}
\min {\bf L}.
\end{equation}

We optimize (19) using the necessary condition for the existence
of an extremum:

$$
\frac{d{\bf L}}{dI_0(l,m)}=0;~~\frac{d{\bf L}}{dI_q^+(l,m)}=0;
~~\frac{d{\bf L}}{dI_q^-(l,m)}=0
$$
for $q=1,...,Q-1$ and $l,m=1,...,N$, and also

$$
\frac{d{\bf L}}{dI\eta^{re}_{u_\nu,v_\nu}}=0;~~\frac{d{\bf
L}}{dI\eta^{im}_{u_\nu,v_\nu}}=0
$$
for all values of the visibility function.

Note that a sufficient condition for the existence of a minimum of
{\bf L} is the positive definiteness of the Hessian matrix.

As a result we obtain the following solution s for the required
unknowns expressed in terms of the dual variables:
\begin{equation}
I_0(l,m)=\exp(-\sum_{u_\nu,v_\nu}[\beta_{u_\nu,v_\nu}
a^{lm}_{u_\nu,v_\nu}+ \gamma_{u_\nu,v_\nu}
b^{lm}_{u_\nu,v_\nu}]-1-\ln a),
\end{equation}
\begin{equation}
I_q^+(l,m)=\exp(-\sum_{u_\nu,v_\nu}[\beta_{u_\nu,v_\nu}
a^{lm}_{u_\nu,v_\nu}+ \gamma_{u_\nu,v_\nu}
b^{lm}_{u_\nu,v_\nu}]\left(\frac{\nu-\nu_0}{\nu_0}\right)^q-1-\ln
a),
\end{equation}
\begin{equation}
I_q^-(l,m)=\exp(\sum_{u_\nu,v_\nu}[\beta_{u_\nu,v_\nu}
a^{lm}_{u_\nu,v_\nu}+ \gamma_{u_\nu,v_\nu}
b^{lm}_{u_\nu,v_\nu}]\left(\frac{\nu-\nu_0}{\nu_0}\right)^q-1-\ln
a),
\end{equation}
\begin{equation}
\eta_{u_\nu,v_\nu}^{re} = -\frac{\sigma_{u_\nu,v_\nu}^2
\beta_{u_\nu,v_\nu}}{\rho}, ~~~~\eta_{u_\nu,v_\nu}^{im} =
-\frac{\sigma_{u_\nu,v_\nu}^2 \gamma_{u_\nu,v_\nu}}{\rho}.
\end{equation}

The solutions (20)–(22) are exclusively positive. To obtain
spectral maps, we must use (12).

It follows from (21) and (22) that the positive and negative parts
of the spectral maps are related as

\begin{equation}
I_q^+(l,m)I_q^-(l,m)=\exp(-2-2\ln a)=K(a),
\end{equation}
with, as was said above, the parameter $a$ playing the role of a
separator between the positive and negative parts of the
$I_q(l,m)$ solution. The higher the value of $a$, the more exact
the separation. The upper limit of $a$ is determined by
computational effects.

It can be readily shown that the Hessian matrix is positively
defined everywhere; therefore, the functional ${\bf L}$ is convex
and the solution is global, i.e., unique.

Substituting the solutions (20)–(23) in to the right-hand side of
(18), we obtain the expression for the dual functional
\begin{equation}
{\bf L}
=\sum_{l,m}I_0(l,m)+\sum_{q=1}^{Q-1}\sum_{l,m}[I_q^+(l,m)+I_q^-(l,m)]+\sum_{u_\nu,v_\nu}[\beta_{u_\nu,v_\nu}A_{u_\nu,v_\nu}+\gamma_{u_\nu,v_\nu}B_{u_\nu,v_\nu}],
\end{equation}
Minimizing this expression [with regard to (20)–(22)] enables us
to find the dual variables $\beta_{u_\nu,v_\nu}$ and
$\gamma_{u_\nu,v_\nu}$.

We can find the extremum of (25) using various gradient methods.We
utilized the coordinate-descent method, considering it to be the
most reliable, though not the fastest. In this method, the
increment of $z_x$ for each unknown
($\beta_{u_\nu,v_\nu},\gamma_{u_\nu,v_\nu}$), which we will
conditionally denote $x$, is searched for based on the condition
for an extremum of ${\bf L}(z_x)$:

\begin{equation}
 \frac{d{\bf
L}(z_x)}{dz_x}=0.
\end{equation}
${\bf L}(z_x)$ is obtained by replacing the variable $x$ in the
functional ${\bf L}$ (26) with $x+z_x$.

In our case, (25) is nonlinear in $z_x$. Obtaining a numerical
solution using the Newton method is an iterative process. At the
$i$th iteration, the value of $z_x$ is
$$
 z_x^i=z_x^{i-1}
 -\Big({d{\bf
L}(z_x)}/{dz_x}\Big)
 /\Big({d^2 {\bf
L}(z_x)}/{dz_x^2}\Big)\Big|_{z_x=z_x^{i-1}}.
$$
The solution for $z_x$ usually converges after one to three
iterations. As a result of one descent, the required variable
acquires a new value, $x+z_x$. Simulation have shown that a
high-quality reconstruction of the radio image and spectral maps
requires from 100 to 500 descents in all dual coordinates
$\beta_{u_\nu,v_\nu},\gamma_{u_\nu,v_\nu}$.

\bigskip

\centerline{5.~SIMULATION RESULTS}

\bigskip

In this section, we present the results of our numerical modeling
of MFS using the above reconstruction based on the generalized
MEM.

As a model radio source at frequency  $\nu_0$, we consider the
intensity distribution $I_0(x,y)$ for two Gaussian components with
amplitudes of 1 and 0.4 Jy/beam (Fig. 2a). The source size is
about 3– 4 mas in both right ascension and declination. Thus, the
source has a size appropriate for VLBI mapping. The model
spectral-index distribution of the source $\alpha(x,y)$ is shown
in Fig. 2c.

The model spectral indices have values from 0 to 0.84. The
distribution of the first-order spectral map,
$I_1(x,y)=I_0(x,y)\alpha(x,y)$, is shown in Fig. 2b. According to
(4), odd-order spectral maps are positive and even-order spectral
maps are negative. The sign of the model spectral-index
distribution is not of primary importance from the viewpoint of
testing the algorithm, since the solutions for the spectral maps
of any order are searched for in the form of sign-variable
functions [see (13) and (25)], and obtaining either positive or
negative solutions is equally valid.

The contour levels in all images in Figs. 2–6 are 0.0625, 0.125,
0.25, 0.5, 1, 2, 4, 8, 16, 32, 64, and 99\% of the peak value;
this allows us to adequately display maps with dynamic ranges of
up to 3200.

The visibility data are formed from the Fourier transform of the
source intensity distribution calculated at each point (x, y) in
accordance with the spectral dependence (1), with measurement
errors added in accordance with (8) and (9).

We adopted $\nu_0=8.2$ GHz as the central, or reference, frequency
of the observations.

To study the quality of the synthesized images as a function of
the range of frequencies used, we carried out simulations for
three cases, with bandwidths of $\pm30\%$, $\pm60\%$, and
$\pm90\%$ of the central frequency. In each case, we used nine
equidistant frequencies. The corresponding UV plane coverages are
shown in Fig. 1.

The primary goal of our simulations is to analyze the quality of
the image reconstruction using various degrees of spectral
corrections, and to establish the minimum necessary number of
spectral terms that must be included in the Taylor series (3) as a
function of the frequency band used. We carried out this analysis
first for fairly accurately measured visibility data (with a
signal-to-noise ratio of about 100) and then for data containing
appreciable measurement errors having a normal distribution with
zero mean and a dispersion yielding a signal-to-noise ratio of
about 10. We chose fairly broad frequency ranges (up to $\pm90\%$
of the central frequency) in order to more fully study the
properties and capabilities of the developed frequency-correction
algorithm, which is of considerable theoretical interest.

As was said in the Introduction, the idea of MFS arises in
connection with the need for fast aperture image synthesis, and is
most urgent for VLBI arrays with relatively few baselines (Fig.
1a). Figure 3a shows an image reconstructed from visibility data
corresponding to the single-frequency coverage of the UV plane
shown in Fig. 1a. Comparing the result of the reconstruction with
the original image, we note severe distortions of the source
shape, as well as some spurious features, leading to a low dynamic
range for the derived map.

The MFS in a $\pm30\%$ band (Fig. 1b) provides a much fuller
coverage of the UV plane. The image reconstructed from the data
corresponding to this coverage (provided the source brightness
does not depend on the frequency) has a substantially higher
quality, with a signal-to-noise ratio of more than 50 (Fig. 3b).

As said above, the MFS is strongly complicated by the spectral
dependence of the source brightness. If the spectral-index
distribution corresponds to the model distribution shown in Fig.
2c, but we have ignored this in the reconstruction, we obtain the
image in Fig. 3c, which has an even poorer quality than the
single-frequency image. Indeed, the resulting image has a
signal-to-noise ratio a factor of three lower than the image in
Fig. 3a. Thus, this simple experiment demonstrates that applying
MFS without taking into account the spectral dependence of the
source emission can result in a deterioration of the
reconstruction.

Our next simulations were directly related to establishing the
relationship between the MFS bandwidth and the minimum necessary
number of terms in the Taylor series expansion (3) of the spectral
component of the image when frequency corrections are applied. It
is clear that, the wider the frequency band, the stronger the
effect of the spectral component and the greater number of terms
in the series required for a satisfactory correction of the
images. However, it is not reasonable to include a very large
number of terms in the expansion, since this results in an
unjustifiably large number of unknowns in the optimization
problem, which can, in turn, lead to a deterioration of the
reconstruction.

Figure 4 shows the results of reconstructing the source emission
distribution, first-order spectral map, and spectral-index
distribution with a frequency band of $\pm30\%$ of the central
frequency (UV coverage of Fig. 1b). According to (6), the
spectral-index distribution is defined by the pixel-by-pixel
division [9] of the reconstructed first-order spectral term
$I_1(x,y)$  by the reconstructed brightness $I_0(x,y)$, for all
pixel values exceeding some empirically chosen threshold, to avoid
division by zero or very small values. The spectral-index maps
were all obtained with the same threshold value of 0.004 of the
peak of $I_0(x,y)$. The upper row of maps were obtained using the
first two terms in the Taylor series expansion (2), and the lower
row of maps using three terms in the expansion; these two sets of
results virtually coincide. A fairly high quality of the
reconstruction has been achieved. Thus, the first-order spectral
term is sufficient when using a fairly narrow frequency band ($\pm
30\%$), consistent with the results of other studies [3--6].

Figure 5 presents the reconstruction results for a $\pm60\%$
frequency band ($UV$ coverage in Fig. 1c). The upper, middle, and
lower rows of the maps were obtained using the first two, three,
and four terms in expansion (3). Using the first-order spectral
term alone does not yield a satisfactory quality of the
reconstruction, while using the third-order spectral term is
unnecessary, since the results using the second term along and
both the second and third terms virtually coincide. Thus, using a
frequency band of up to $60\%$ requires including higher-order
terms together with the first-order spectral term. Addition of the
second-order spectral term approximately doubles the resulting
signal-to-noise ratio.

Figure 6 shows the reconstruction results for a much broader
frequency band reaching $\pm90\%$ of the central frequency ($UV$
coverage in Fig. 1d). The rows of maps from top to bottom were
obtained using the first two, three, four, and five terms in
expansion (3). Including only the first-order spectral term in the
spectral correction is clearly not sufficient. Including the
second-order spectral term appreciably improves the quality of the
reconstruction, increasing the signal-to-noise ratio by almost a
factor of two. Including the third-order spectral term provides a
further substantial improvement in the quality of the
reconstruction, increasing the signal-to-noise ratio by another
factor of 1.5. Including the fourth-order spectral term results in
a further, though not so significant, improvement in the quality
of the reconstruction. Thus, using a frequency band of up to 90\%
requires taking into account spectral terms at least up to third
order.

The results of our simulations demonstrate that the highest
accuracy of the reconstruction can be achieved for the $I_0(x,y)$
source images, while the maps of the first-order spectral term
$I_1(x,y)$ have a lower accuracy, and the spectral-index maps
$\alpha(x,y)$ have the lowest accuracy. In particular, the maximum
signal-to-noise ratios achieved for the $I_0(x,y)$, $I_1(x,y)$,
and $\alpha(x,y)$ maps are 40, 15, and 4, respectively

An intercomparison of the best reconstruction results obtained
using each frequency band demonstrates that, for the same number
of frequencies within the band, the quality of the reconstruction
degrades slightly with increasing bandwidth. This is true because,
with increasing frequency bandwidth, we must increase the number
of spectral terms taken into account. In turn, for the same number
of visibility measurements, the number of unknowns in the
optimization of (25) is increased. To improve the reconstruction
quality of images with increasing frequency bandwidths, we must
proportionally increase the number of frequencies in the band. The
simulations demonstrate (Fig. 7) that using 27 (instead of 9)
frequencies in a $\pm90\%$ band increases the signal-to- noise
ratio to 54 and the image dynamic range to 3200.

Figure 8 shows the reconstruction results obtained from data
containing appreciable errors. The dynamic range of the maps
obtained is reduced by approximately a factor of four due to the
effect of the errors. The contour levels on these images are 0.25,
0.5, 1, 2, 4, 8, 16, 32, 64, and 99\% of the peak value. Figure 8a
presents the radio map reconstructed from single-frequency data
(Fig. 1a), which displays a higher level of distortion than the
analogous map in Fig. 3a. Figure 8b shows the MFS map obtained
using a $\pm90\%$ band with nine frequencies (Fig. 1d), neglecting
the spectral dependence of the source emission. This map is
likewise characterized by strong distortions, and its
signal-to-noise ratio is only 1.4, a factor of six lower than the
previous map. Figures 8c, 8d, 8e, and 8f present MFS maps
reconstructed from the same data including two, three, four, and
five terms in the expansion (2), respectively. Adding subsequent
terms to the expansion substantially improves the quality of the
reconstruction. The inclusion of spectral terms up to fourth order
improves the signal-to-noise ratio to 20, twice the
signal-to-noise ratio in the data. Figures 8g and 8h show the maps
of the first-order spectral term and the spectral index
reconstructed using five terms in expansion (2). The
spectral-index map was obtained using(7) with a threshold value of
0.01 of the peak in the $I_0(x,y)$ map. The signal-to-noise ratios
of the spectral-term and spectral-index maps are six and two,
respectively. A comparison of these images with the model maps
indicates a satisfactory reconstruction of the source intensity
(the dynamic range is about 400) and spectral-index distributions,
even in the presence of a considerable noise in the data. The
quality of the reconstruction is substantially improved by
increasing the number of frequencies within the frequency band. In
particular, if the number of frequencies is tripled, the dynamic
range of the maps is approximately doubled.

\bigskip

\centerline{6.~CONCLUSION}

\bigskip

Multifrequency synthesis is an effective means of synthesizing the
aperture of a radio interferometer, which can enable
high-dynamic-range imaging of radio sources. The main problem that
must be solved in MFS is the need to apply a frequency correction
to the images, due to the frequency dependence of the source
brightness. Deconvolution algorithms incorporating such a
frequency correction based on the CLEAN method are well known. The
best studied of these is a double-deconvolution algorithm intended
for linear spectral correction. This algorithm is effective when
applied to frequency bandwidths no broader than $\pm30\%$.

We have proposed, developed, and studied a more fundamental
deconvolution method for MFS images based on a generalized maximum
entropy method. This new method makes it possible to apply
frequency corrections in a much broader frequency band, due to the
ability to include non-linear terms of arbitrary order, together
with the linear term of the spectral component of the image. The
main advantage of MEM over CLEAN -- the ability to more accurately
restore extended sources -- is also realized. In addition, the
developed algorithm enables acceptably accurate estimation of the
spectral-index distribution.

Our numerical simulations have enabled us to establish
regularities in the quality of the image reconstruction associated
with changes in the frequency bandwidth, number of frequencies
within the frequency band, number of spectral terms included, and
the noise level in the input data. We conclude that our developed
deconvolution method with frequency correction is most effective
for mapping extended sources in a broad frequency band using
high-precision visibility data. The proposed spectral correction
method can be further developed using the principle of difference
mapping [19], intended to improve the quality of image
reconstruction for sources whose structure includes both compact
and extended components. Another subject of further studies is a
combination of MFS with phaseless mapping [20], including also
data from a high-orbit space radio interferometer [21].

\bigskip

\centerline{ACKNOWLEDGMENTS}

\bigskip

This work was supported by the Basic-Research Program of the
Presidium of the Russian Academy of Sciences on "The Origin and
Evolution of Stars and Galaxies" and the Program of State Support
for Leading Scientific Schools of the Russian Federation (grant
NSH-6110.2008.2, "Multi-wavelength Astrophysical Research"). The
author thanks the referee for valuable comments, which helped to
improve the paper.

\bigskip

\centerline{REFERENCES}

\bigskip

1. A. R. Thompson, J. M. Moran, and G. W. Swenson, Interferometry
and Synthesis in Radio Astronomy, (Wiley, New York, 1986;
Fizmatlit, Moscow, 2003).

2. A. T. Bajkova and A. M. Finkelstein, Astron. Astrophys. Trans.
{\bf 1}, 159 (1992).

3. J. E. Conway, T. J. Cornwell, and P. N. Wilkinson, Mon. Not. R.
Astron. Soc. {\bf 246}, 490 (1990).

4. J. E. Conway, ASP Conf. Ser. {\bf 19}, 171 (1991).

5. R. J. Sault and M. H. Wieringa, Astron. Astrophys. Suppl. Ser.
{\bf 108}, 585 (1994).

6. R. J. Sault and J. E. Conway, in: Synthesis imaging in Radio
Astronomy II, eds G. B. Taylor,C. L. Carilli, R. A. Perley, ASP
Conf. Ser. {\bf 180}, 419 (1999).

7. R. J. Sault and T. A. Oosterloo, e-Print arXiv:astroph/
0701171v1 (2007).

8. J. A. H\"{o}gbom, Astron. Astroph. Suppl. Ser. {\bf 15}, 417
(1974).

9. S. F. Likhachev, V. A. Ladynin, and I. A. Girin, Izv.
Vyssh.Uchebn. Zaved., Ser.Radiofiz. {\bf 49}, 553 (2006).

10. B. R. Frieden, J. Opt. Soc. Amer. {\bf 62}, 511 (1972).

11. J. G. Ables, Astron. Astrophys. Suppl. Ser. {\bf 15}, 383
(1974).

12. J. Skilling and R.K.Bryan, Mon.Not. R. Astron. Soc. {\bf 211},
111 (1984).

13. S. J. Wernecke and L. R. D'Addario, IEEE Trans. Comp. {\bf
C-26}, 351 (1977).

14. T. J. Cornwell and K. F. Evans, Astron. Astrophys. {\bf 143},
77 (1985).

15. R. Narayan and R. Nityananda, Ann. Rev. Astron. Astrophys.
{\bf 24}, 127 (1986).

16. T. J. Cornwell, R. Braun, and D. S. Briggs, in: Synthesis
imaging in Radio Astronomy II, eds G. B. Taylor, C. L. Carilli, R.
A. Perley, ASP Conf. Ser. {\bf 180}, 151 (1999).

17. A. T. Bajkova, Rep. Institute Appl.Astron.No. 58 (Inst. Appl.
Astron. Ross. Akad. Nauk, S.-Petersburg, 1993).

18. B. R. Frieden and A. T. Bajkova, Appl. Opt. {\bf 33}, 219
(1994).

19. A. T. Bajkova, Astron. Zh. {\bf 84}, 984 (2007) [Astron. Rep.
51, 891 (2007)].

20. A. T. Bajkova, Pis'ma Astron. Zh. 30, 253 (2004) [Astron.
Lett. {\bf 30}, 218 (2004)].

21. A. T. Bajkova, Astron. Zh. {\bf 82}, 1087 (2005) [Astron. Rep.
49, 947 (2005)].

\newpage

\newpage
\begin{figure}[t]
{\begin{center}
 \includegraphics[width= 100mm]{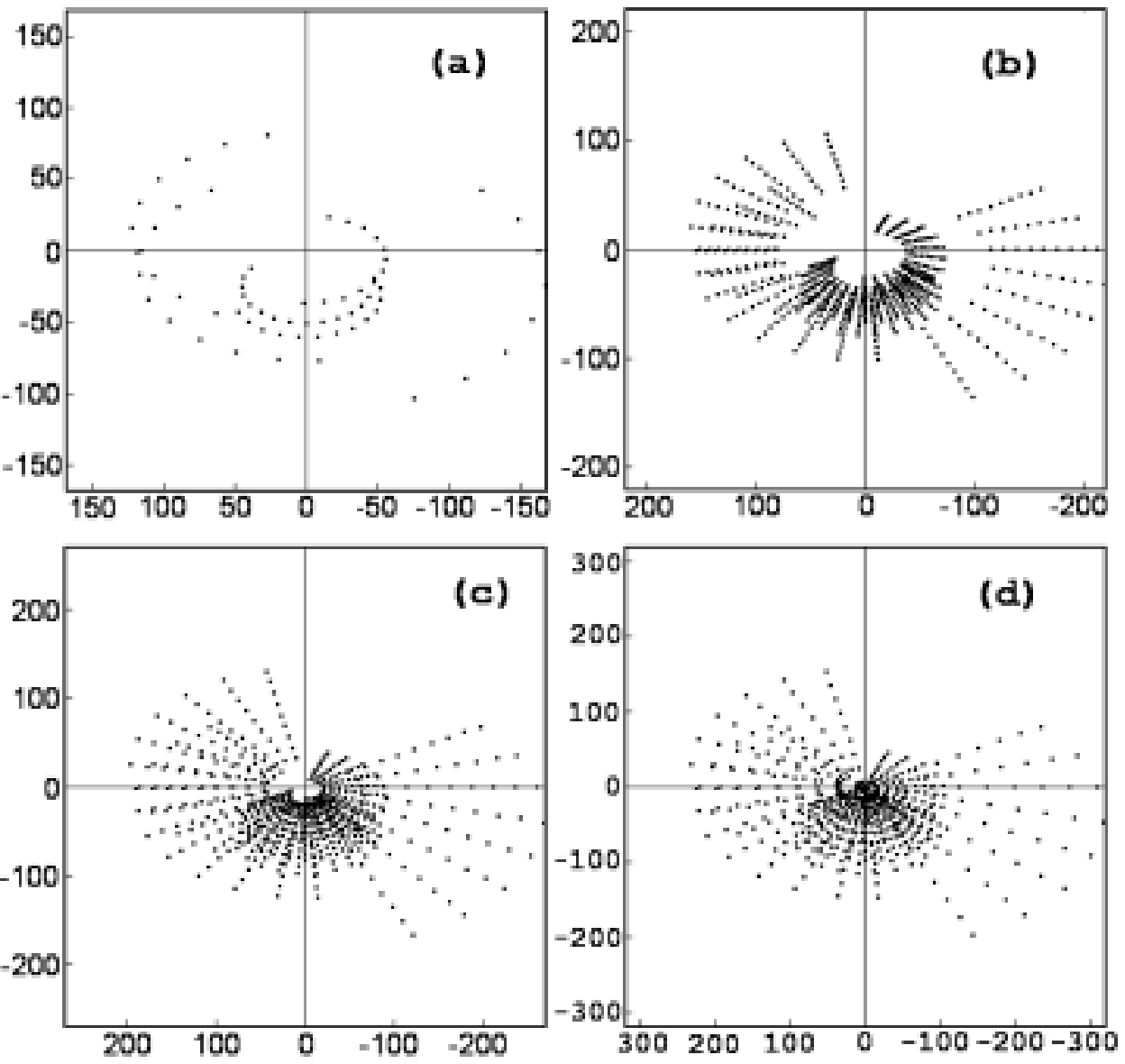}
\end{center}
Fig. 1. $UV$ coverages for a four-element radio interferometer
comprised of the Svetloe, Zelenchukskaya, Badary, and Matera
stations for a 12-hour observation of a radio source with a
declination of $34^o$, for (a) single-frequency observations
($\nu_0=8.2$ GHz) and multi-frequency synthesis in a band
encompassing(b) $\pm30\%, \pm60\% $, and (d) $\pm90\%$ of the
reference frequency (nine frequency channels are used in each
band). The horizontal and vertical axes show the $U$ and $V$
spatial frequencies in millions of wavelengths. }
\end{figure}

\newpage
\begin{figure}[t]
{\begin{center}
 \includegraphics[width= 160mm]{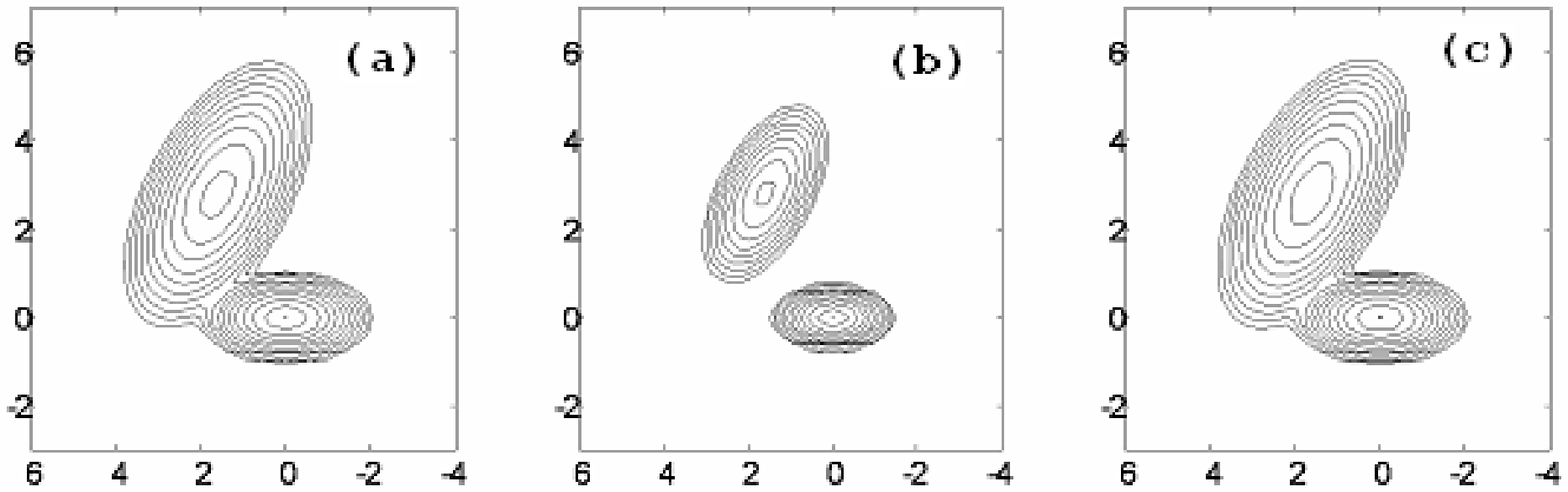}
\end{center}
Fig. 2. Model (preset) (a) radio intensity map $I_0(x,y)$, (b)
first-order spectral map $I_1(x,y)$, and (c) spectral-index map
$\alpha(x,y)$. The horizontal and vertical axes plot right
ascension $x$ and declination $y$ in milliarcseconds.

}

\end{figure}

\newpage
\begin{figure}[t]
{\begin{center}
 \includegraphics[width= 160mm]{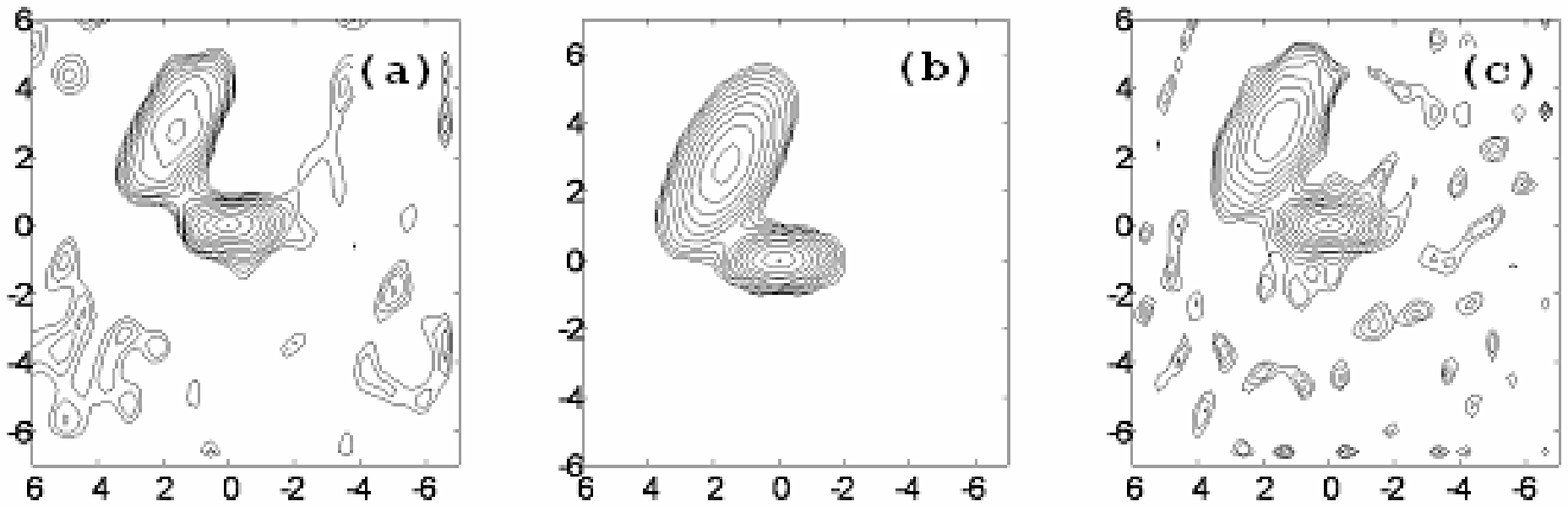}
\end{center}
Fig. 3. Restored maps of the radio source using( a)
single-frequency data, (b) MFS data in a $\pm30\%$ band assuming
frequency independence of the image ($\alpha(x,y)=0$), and (c)
from MFS data in the same band without assuming frequency
dependence of the image ($\alpha(x,y)\ne 0$). }
\end{figure}

\newpage
\begin{figure}[t]
{\begin{center}
 \includegraphics[width= 160mm]{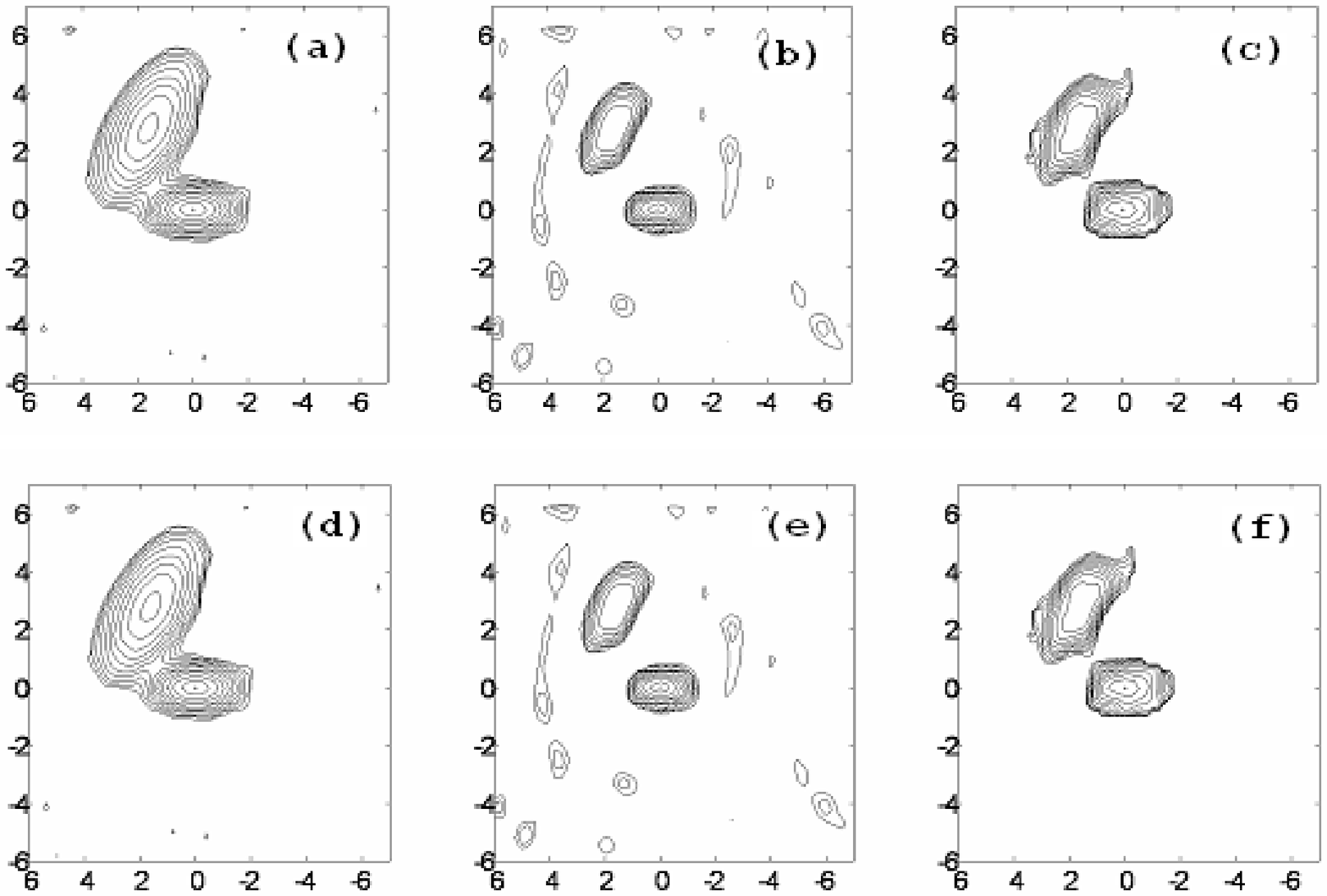}
\end{center}
Fig. 4. Results of the reconstruction using MFS data in a
$\pm30\%$ band. The horizontal ordering of the maps is as in Fig.
2. Maps of the upper row (a), (b), (c) were obtained using the
first two terms, and maps of the lower row (d), (e), (f) the first
three terms of the Taylor series expansion. }
\end{figure}

\newpage
\begin{figure}[t]
{\begin{center}
 \includegraphics[width= 160mm]{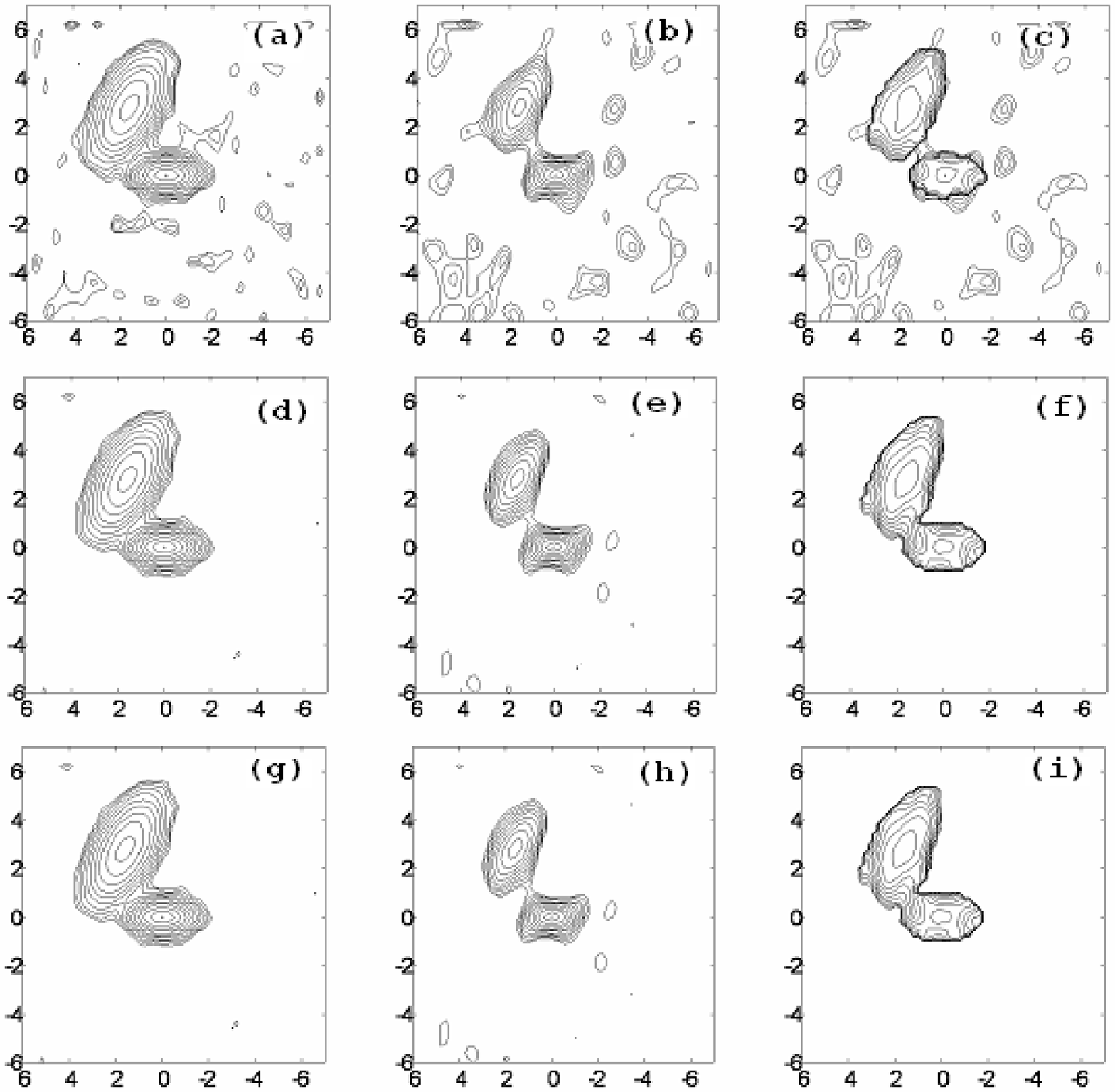}
\end{center}
Fig. 5. Results of the reconstruction using MFS data in a
$\pm60\%$ band. The horizontal ordering of the maps is as in Fig.
2. Maps of the upper row (a), (b), (c) were obtained using the
first two terms, maps of the middle row (d), (e), (f) the first
three terms, and maps of the lower row (g), (h), (i) the first
four terms of the Taylor series expansion. }
\end{figure}

\newpage
\begin{figure}[t]
{\begin{center}
 \includegraphics[width= 160mm]{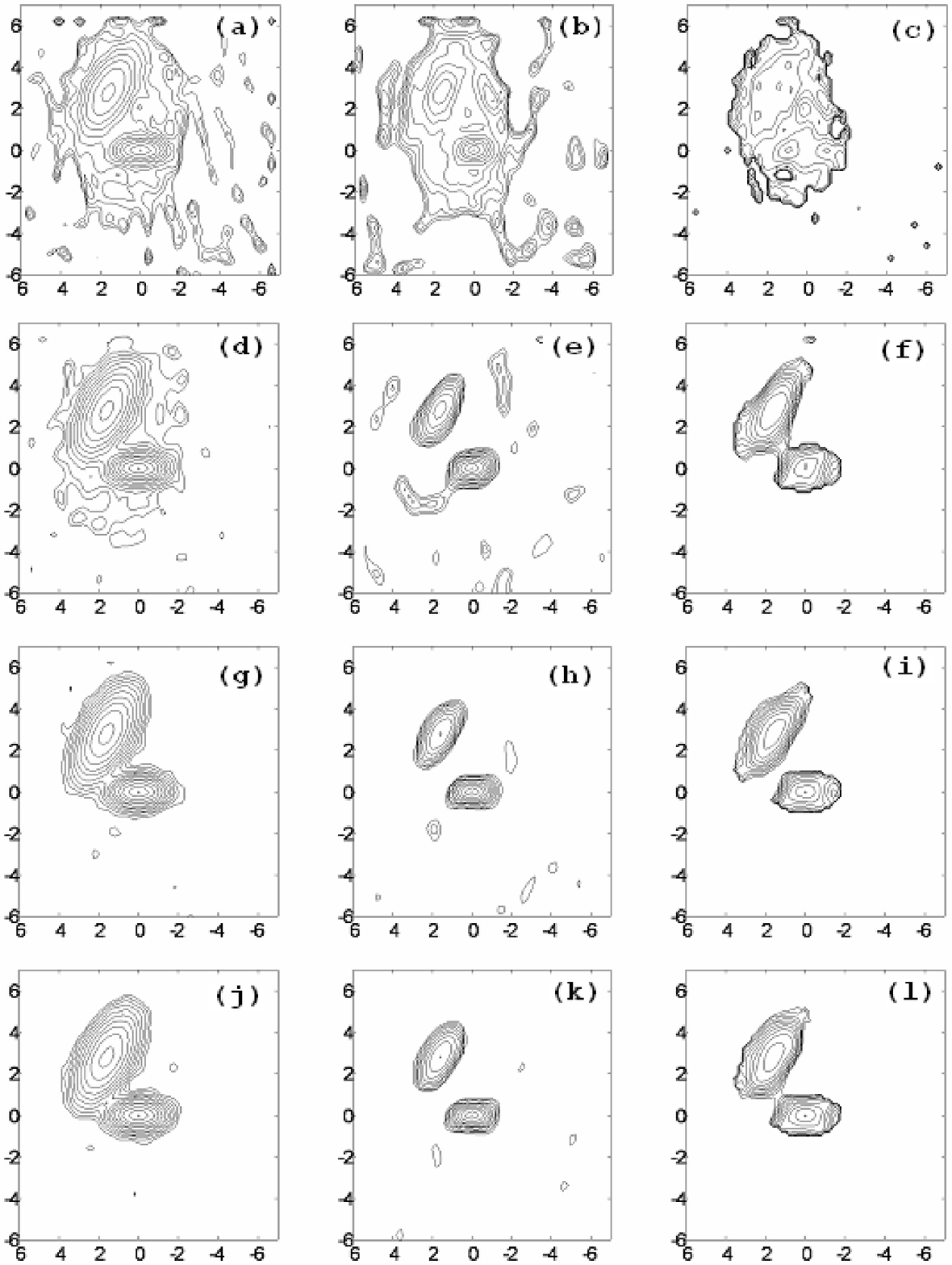}
\end{center}
Fig. 6. Results of the reconstruction using MFS data in a
$\pm90\%$ band. The horizontal ordering of the maps is as in Fig.
2. Maps of the first row (a), (b), (c) were obtained using the
first two terms, maps of the second row (d), (e), (f) the first
three terms, maps of the third row (g), (h), (i) the first four
terms, and maps of the fourth row (j), (k), (l) the first five
terms of the Taylor series expansion. }
\end{figure}

\newpage
\begin{figure}[t]
{\begin{center}
 \includegraphics[width= 160mm]{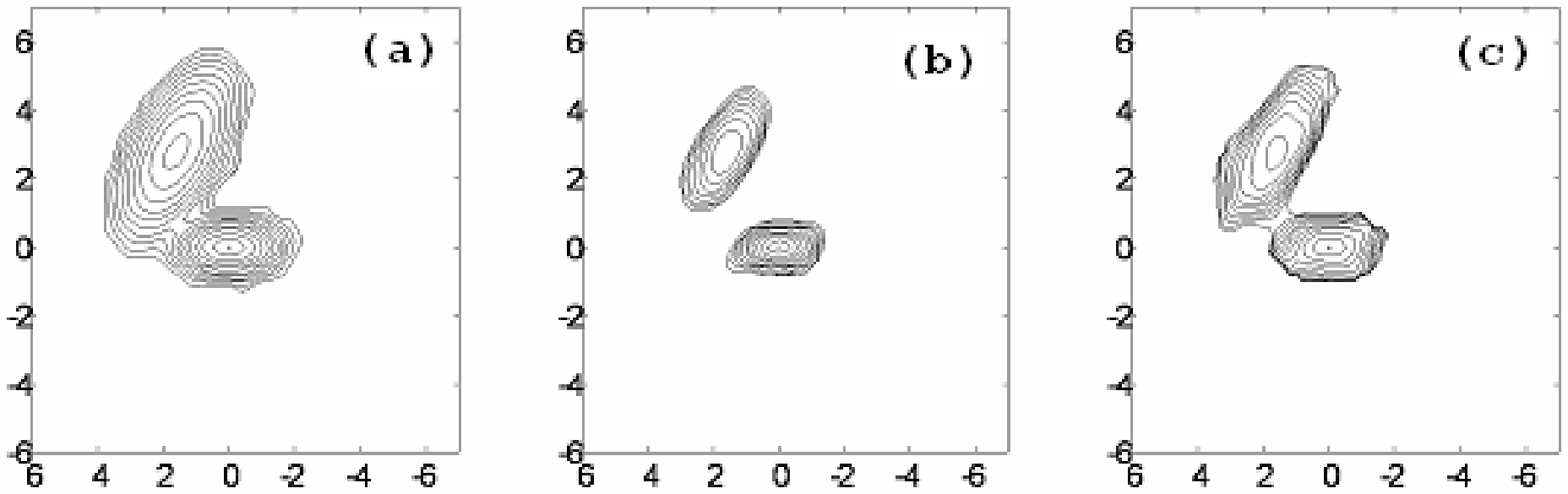}
\end{center}
Fig. 7. Results of the reconstruction using MFS data in a
$\pm90\%$  band with 27 frequency channels using five terms in the
Taylor series expansion. The horizontal ordering of the maps is as
in Fig. 2. }
\end{figure}

\newpage
\begin{figure}[t]
{\begin{center}
 \includegraphics[width= 100mm]{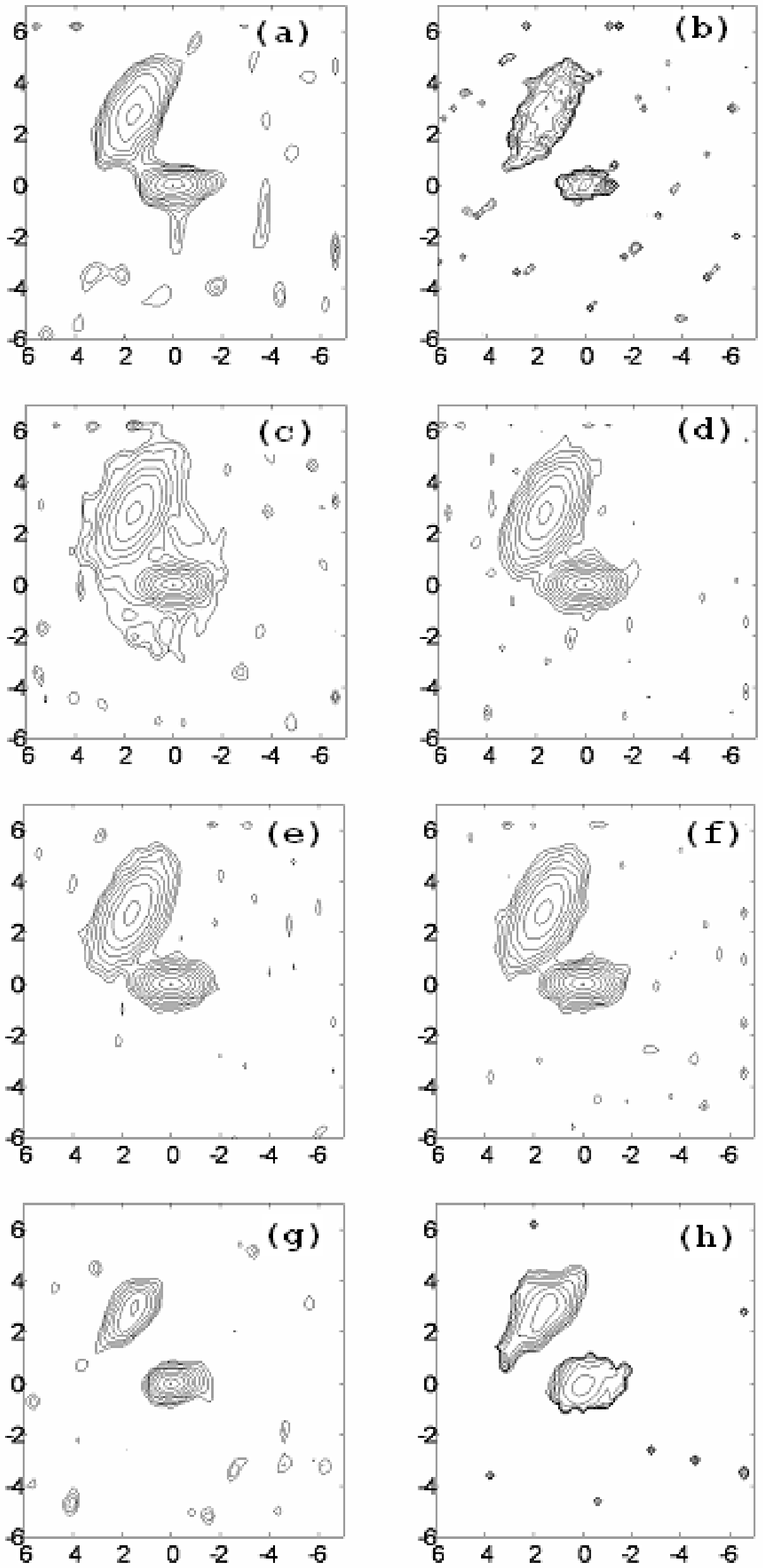}
\end{center}
Fig. 8. Results of the reconstruction with appreciable data errors
(a) for single-frequency data (UV coverage of Fig. 1a); (b) for
MFS with a $\pm90\%$ band without assuming frequency dependence of
the map. Panels (c), (d), (e) and (f) show the radio images
obtained using two, three, four, and five terms of the Taylor
series expansion of the spectral term, respectively, and panels
(g) and (h) the first-order spectral map and spectral-index map
corresponding to map (f). }
\end{figure}

\end{document}